\documentclass[preprint,11pt]{elsarticle}
\usepackage{lscape}
\usepackage{graphicx}
\usepackage{upgreek}
\usepackage{caption} 
\usepackage{hyperref}
\usepackage{todonotes}
\usepackage{amssymb}
\usepackage{url}
\usepackage{array}
\usepackage{enumitem}
\usepackage{soul}
\usepackage{color}
\usepackage{pgfplots}
\pgfplotsset{compat=newest,compat/show suggested version=false}
\usepackage{multirow}
\usepackage{pbox}
\usepackage{float}
\usepackage{longtable}
\usepackage[normalem]{ulem}
\usepackage[export]{adjustbox}
\usepackage{colortbl}

\definecolor{gray50}{gray}{.5}
\definecolor{gray40}{gray}{.6}
\definecolor{gray30}{gray}{.7}
\definecolor{gray20}{gray}{.8}
\definecolor{gray10}{gray}{.9}
\definecolor{gray05}{gray}{.95}

\usepackage{tikz}
\tikzstyle{mybox} = [draw=black, very thick, rectangle, rounded corners, inner ysep=5pt, inner xsep=5pt]
\definecolor{darkmagenta}{rgb}{0.55, 0.0, 0.55}

\tikzstyle{mybox} = [draw=black, very thick, rectangle, rounded corners, inner ysep=5pt, inner xsep=5pt]
\usepackage{color}
\newcommand\rf[1]{\textcolor{black}{#1}}

\journal{Information and Software Technology}

\begin{document}

\begin{frontmatter}


\title{From Monolithic Systems to Microservices: \\ An Assessment Framework} 



\author[UI]{Florian Auer}
\ead{Florian.Auer@uibk.ac.at}
\author[LUT]{Valentina Lenarduzzi}
\ead{valentina.lenarduzzi@lut.fi}
\author[UI,BTH]{Michael Felderer}
\ead{Michael.Felderer@uibk.ac.at}
\author[TUT]{Davide Taibi}
\ead{davide.taibi@tuni.fi}

\address[UI]{University of Innsbruck, Austria}
\address[LUT]{LUT University, Finland}
\address[BTH]{Blekinge Institute of Technology, Sweden}
\address[TUT]{Tampere University, Finland}

\begin{abstract}
Context. Re-architecting monolithic systems with Microservices-based architecture is a common trend. Various companies are migrating to Microservices for different reasons. However, making such an important decision like re-architecting an entire system must be based on real facts and not only on gut feelings. \\
Objective. The goal of this work is to propose an evidence-based decision support framework for companies that need to migrate to Microservices, based on the analysis of a set of characteristics and metrics they should collect before re-architecting their monolithic system. \\
Method. We conducted a survey done in the form of interviews with professionals to derive the assessment framework based on Grounded Theory. \\
Results. We identified a set consisting of information and metrics that companies can use to decide whether to migrate to Microservices or not. The proposed assessment framework,  based on the aforementioned metrics,  
could be useful for companies if they need to migrate to  Microservices and do not want to run the risk of failing to consider some important information. \\

\end{abstract}

\begin{keyword}
Microservices \sep Cloud Migration \sep Software Measurement


\end{keyword}
\end{frontmatter}


\section{Introduction}
\label{Intro}

Microservices are becoming more and more popular. Big players such as Amazon~\footnote{https://gigaom.com/2011/10/12/419- the-biggest-thing-amazon-got-right-the-platform/}, Netflix~\footnote{http://nginx.com/blog/Microservices-at-netflix-architectural-best- practices/}, Spotify~\footnote{ www.infoq.com/presentations/linkedin-Microservices-urn}, as well as small and medium-sized enterprises are developing Microservices-based systems~\cite{TaibiIEEECloud}. 

Microservices are
autonomous services deployed independently, with a single and clearly defined purpose~\cite{Fowler2014}. 
Microservices propose vertically decomposing applications into a subset of business-driven independent services. Each service can be developed, deployed, and tested independently by different development teams and using different technology stacks. Microservices have a variety of different advantages. They can be developed in different programming languages, can scale independently from other services, and can be deployed on the hardware that best suits their needs. Moreover, because of their size, they are easier to maintain and more fault-tolerant since the failure of one service will not disrupt the whole system, which could happen in a monolithic system.
However, the migration to Microservices 
is not an easy task~\cite{TaibiIEEECloud}~\cite{Soldani2018}. Companies commonly start the migration without any experience with Microservices, only rarely hiring a consultant to support them during the migration~\cite{TaibiIEEECloud}~\cite{Soldani2018}.

Various companies are adopting Microservices since they believe that it will facilitate their software maintenance. In addition, companies hope to improve the delegation of responsibilities among teams. Furthermore, there are still some companies that refactor their applications with a Microservices-based architecture just to follow the current trend~\cite{TaibiIEEECloud}~\cite{Soldani2018}. 

The economic impact of such a change is not negligible, and taking such an important decision to re-architect an existing system should always be based on solid information, so as to ensure that the migration will allow achieving the expected benefits.
 
In this work, we propose an evidence-based decision support framework to allow companies, and especially software architects, to make their decision on migrating monolithic systems to Microservices based on the evaluation of a set of objective measures regarding their systems. The framework supports companies in discussing and analyzing potential benefits and drawbacks of the migration and re-architecting process.

For this purpose we designed and conducted interviews with experienced practitioners as participants, to understand which characteristics and metrics they had considered before the migration and which they should have considered, comparing the usefulness of the collection of these characteristics. Finally, based on the application of Grounded Theory on the interviews, we developed our decision support framework. 


\textbf{Paper structure}. Section~\ref{Background} presents the background and Section~\ref{RW} the related work. 
Section~\ref{Survey} presents the design and the results of the survey. In Section~\ref{Framework}, we present the defined framework. In Section~\ref{Discussion}, we discuss the results we obtained and the defined framework. In Section~\ref{Threats}, we identify threats to the validity of this work. Finally, we draw conclusions in Section~\ref{Conclusion} and highlight future work.

\section{Background}
\label{Background}
 


The Microservice architecture pattern emerged from Service-Oriented Architecture (SOA). Although services in SOA have dedicated responsibilities, too, they are not independent. The services in such an architecture cannot be turned on or off independently. This is because the individual services are neither full-stack (e.g., the same database is shared among multiple services) nor fully autonomous (e.g., service A depends on service B). As a result, services in SOA cannot be deployed independently.

\begin{figure*}[!ht]
\centering 
\includegraphics[trim= 0 50 0 0, clip, width=1\linewidth]{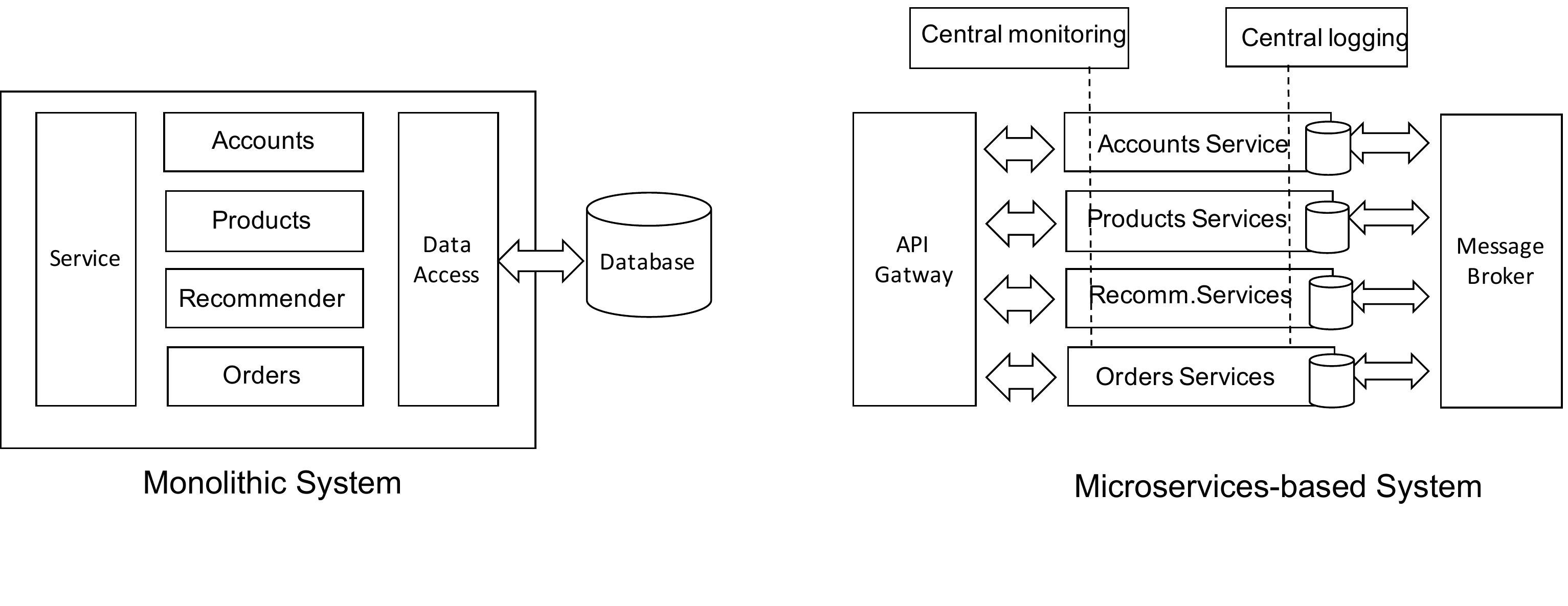}
\caption{Comparison between Microservices and monolithic architectures}
\label{fig:Microservice}
\end{figure*}

In contrast, Microservices are independent, deployable, and have a lot of advantages in terms of continuous delivery compared to SOA services. They can be developed in different programming languages, can scale independently from other services, and can be deployed on the hardware that best suits their needs because of their autonomous characteristics. Moreover, their typically small size, compared to large monolithic systems, facilitates maintainability and improves the fault tolerance of the services. One consequence of this architecture is that the failure of one service will not disrupt the whole system, which could happen in a monolithic system~\cite{Fowler2014}. Nevertheless, the overall system architecture changes dramatically (see Figure \ref{fig:Microservice}). One monolithic service is broken down into several Microservices. Thus, not only the service's internal architecture changes, but also the requirements on the environment. Each Microservice can be considered as a full-stack that requires a full environment (e.g., its own database, its own service interface). Hence, coordination among the services is needed.

\section{Related Work}
\label{RW}
In this section, we analyze the characteristics and measures adopted by previous studies, in order to classify the characteristics and metrics adopted in empirical studies that compared monolithic and Microservices-based systems. 

\subsection{Microservice Migration}
Many studies concerning specific characteristics of them have already been published. However, there are still some challenges in understanding how to develop such kinds of architectures~\cite{Balalaie2016}~\cite{TaibiBOOK}~\cite{TaibiIEEEsw}. A few secondary studies in the field of Microservices (i.e., \cite{Soldani2018}, \cite{Malavolta2017}, \cite{Malavolta2018}, \cite{TaibiBOOK1}, \cite{TaibiCLOSER} and \cite{Pahl16}) have synthesized the research in this field and provide an overview of the state of the art and further research directions.

Di Francesco et al.~\cite{Malavolta2017} studied a large corpus of 71 studies in order to identify the current state of the art on Microservices architecture. They found that the number of publications about Microservices sharply increased in 2015. In addition, they observed that most publications are spread across many publication venues and concluded that the field is rooted in practice.
In their follow-up work, Di Francesco et al.~\cite{Malavolta2018}, provided an improved version, considering 103 papers. 

Pahl et al.~\cite{Pahl16} covered 21 studies. They discovered, among other things, that most papers are about technological reviews, test environments, and use case architectures. Furthermore, they found no large-scale empirical evaluation of Microservices. These observations made them conclude that the field is still immature. Furthermore, they stated a lack of deployment of Microservice examples beyond large corporations like Netflix. 

Soldani et al.~\cite{Soldani2018} identified and provided a taxonomic classification comparing the existing gray literature on the pains and gains of Microservices, from design to development. They considered 51 industrial studies. Based on the results, they prepared a catalog of migration and re-architecting patterns in order to facilitate re-architecting non-cloud-native architectures during migration to a cloud-native Microservices-based architecture. 

All studies agree that it is not clear when companies should migrate to Microservices and which characteristics the companies or the software should have in order to benefit from the advantages of Microservices.

Thus, our work is an attempt to close this gap by providing a set of characteristics and measures together with an assessment framework, as designed in our previous proposal~\cite{Auer2018}.


\subsection{Characteristics and measures investigated in empirical studies on Microservices}
\label{MappingStudy}


Different product and process characteristics and measures have been investigated in the literature while comparing monolithic systems with Microservices architectures. 

Different studies focused only on product characteristics (\cite{SP1}, \cite{SP2}, \cite{SP4}, \cite{SP5}, \cite{SP6}, \cite{SP7}, \cite{SP8}, \cite{SP9}, \cite{SP10}, \cite{SP11}, \cite{SP12}), on process characteristics (\cite{SP1}, \cite{SP3}, \cite{SP9}, \cite{SP11}, \cite{SP12}) or on both (\cite{SP1}, \cite{SP4}, \cite{SP9}, \cite{SP11}, \cite{SP12}, \cite{TaibiSysta2019}, \cite{TaibiSysta2020}, \cite{Lenarduzzi2020}). Moreover, other studies  (\cite{SP1}, \cite{SP3}, \cite{SP5}) investigated and compared costs. Furthermore, other studies, investigated several characteristics at the same time~\cite{SP1}.

As for the product characteristics,
the most frequently addressed one is performance (see Table \ref{tab:ProcessBenefits}). In detail, the papers \cite{SP1}, \cite{SP2}, \cite{SP4}, \cite{SP5}, \cite{SP6}, \cite{SP7}, \cite{SP8}, \cite{SP9}, \cite{SP11} have a focus on performance. This is followed by scalability, which is discussed by the papers \cite{SP2},\cite{SP4},\cite{SP5},\cite{SP6},\cite{SP7},\cite{SP8}, \cite{SP10}, and \cite{SP11}.
Other characteristics like availability (\cite{SP4}, \cite{SP9}) or maintenance (\cite{SP1},\cite{SP5}, \cite{SP7}, \cite{SP12}) are considered only in a few papers. 

Overall, related works identified the following characteristics as reported in Tables \ref{tab:ProductBenefits}, \ref{tab:ProcessBenefits}, and \ref{tab:CostComparison}:
\begin{itemize}
\item \textbf{Product}
\begin{itemize}
\item Performance
\item Scalability
\item Availability
\item Maintenance
\end{itemize}
\item \textbf{Process}
\item \textbf{Cost}
\begin{itemize}
\item Personnel Cost
\item Infrastructure Cost
\end{itemize}
\end{itemize}

From the literature, we also identified 18 measures for measuring product process and cost, as reported in Tables~\ref{tab:ProductBenefits}, \ref{tab:ProcessBenefits}, and~\ref{tab:CostComparison}. 


\textbf{Product-related measures}. 
We identified 13 measures (Table~\ref{tab:ProductBenefits}) for the four identified sub-characteristics (performance, scalability, availability, and maintenance).

From the obtained results, we can see that the highest number of measures is related to performance and scalability, where we identified a total of nine studies referring to them. Among them, response time, number of requests per minute or second, and waiting time are the most commonly addressed measures. 
For availability, we derived only three measures and for maintainability only two.

\textbf{Process-related measures}. Seven studies investigated the migration process using three factors: development independence between teams, usage of continuous delivery, and reusability (Table~\ref{tab:ProcessBenefits}). These three factors can be considered as ''Boolean measures'' and can be used by companies to understand whether their process can be easily adapted to the development of Microservices-based systems. 

Existing independent teams could easily migrate and benefit from the independence freedom provided by Microservices. Continuous delivery is a must in Microservices-based systems. The lack of a continuous delivery pipeline eliminates most of the benefits of Microservices. Reusability is amplified in Microservices. Therefore, systems that need to reuse the same business processes can benefit more from Microservices, while monolithic systems in which there is no need to reuse the same processes will not experience the same benefits. 

Besides the analyzed characteristics, the papers also discuss several process-related benefits of the migration. Technological heterogeneity, scalability, continuous delivery support, and simplified maintenance are the most frequently mentioned benefits. Furthermore, the need for recruiting highly skilled developers and software architects is considered as a main motivation for migrating to Microservices.

\textbf{Cost-comparison-related measures}. 
As for this characteristic, three studies include it in their analysis and consider three measures for the comparison (Table~\ref{tab:CostComparison}). 

\begin{table}[H]
\caption{Product-related measures}
\label{tab:ProductBenefits}
\centering
\scriptsize
\begin{tabular}
{@{}p{2cm}|p{10cm}@{}}
\hline 
\textbf{Characteristic} & \textbf{Measures}\\ \hline 
\multirow{5}{*}{\pbox{2.2cm}{Performance}} 
& \textbf{Response time:} The time between sending a request and receiving the corresponding response. This is a common metric for measuring the performance impact of approaches (\cite{SP1}\rf{,} \cite{SP4}, \cite{SP5}, \cite{SP6}, \cite{SP8}, \cite{SP9}, \cite{SP11}).\\
& \textbf{CPU utilization:} The percentage of time the CPU is not idle. Used to measure performance. \cite{SP9} reports the relationship between the number of VMs and the overall VMs utilization. In addition, \cite{SP11} analyzes the impact of the decision between VMs and containers on CPU utilization. \\
& \textbf{Impact of programming language:} \rf{Communication between Microservices is network-based. Hence, network input and output operations require a considerable amount of the total processing time. The network performance is influenced amongst others by the selection of the programming language. That is due to the different implementations of the communication protocols. \cite{SP7} analyzed the impact of design decisions on the performance and recommend specific programming languages for specific ranges of network message sizes. However, considering scalability in the system design seems to mitigate programming language impact. An example therefore is the routing mechanism for Microservices proposed in \cite{SP8}.} \\
& \textbf{Path length:} The number of CPU instructions to process a client request. \cite{SP2} reports that the length of the code path of a Microservice application developed using Java with a hardware configuration of one core, using a bare process, docker host, and docker bridge, is nearly twice as high as in a monolithic system. \\
& \textbf{Usage of containers:} The usage of containers can influence performance, since they need additional computational time compared to monolithic applications deployed in a single container. \cite{SP7} reports that the impact of containers on performance might not always be negligible. \\
& \rf{\textbf{Waiting time:} The time a service request spends in a waiting queue before it gets processed. \cite{SP6}, \cite{SP10} discuss the relationship between waiting time and number of services. Furthermore, \cite{SP8} mentions an architecture design that halves the waiting time compared to other design scenarios.}\\
\hline 
\multirow{3}{*}{\pbox{2.2cm}{Scalability}} 
& \textbf{Number of requests per minute or second:} (also referred to as throughput \cite{SP2,SP5,SP11} or average latency \cite{SP4,SP7}), is a performance metric. \cite{SP11} found that in their experimental setting, the container-based scenario could perform more requests per second than the VM-based scenario. \\
& \textbf{Number of features per Microservice:} \cite{SP10} points out that the number of features per Microservice affects scalability, influences communication overhead, and impacts performance.\\
\hline 
\multirow{3}{*}{Availability} 
& \textbf{Downtime:} \rf{Microservice might suffer of downtime, if the system is not properly designed}~\cite{SP4}\cite{Saquicela21}\cite{marquez20}. 
\\
& \textbf{Mean time to recover:} The mean time it takes to repair a failure and return back to operations. \cite{SP9} uses this measure to quantify availability. \\
& \textbf{Mean time to failure:} The mean time until the first failure. \cite{SP9} uses this measure together with mean time to recover as a proxy for availability. \\
\hline 
\multirow{3}{*}{Maintenance} 
& \textbf{Complexity:} \cite{SP1}, \cite{SP5} notes that Microservices reduce the complexity of a monolithic application by breaking it down into a set of services. However, some development activities like testing may become more complex \cite{SP5}. Furthermore, \cite{SP7} state that the usage of different languages for different Microservices increases the overall complexity. \\
& \textbf{Testability:} \cite{SP12} concludes that the loose coupling of Microservices at the application's front-end level improves testability. \\ \hline 
\end{tabular}
\end{table}

\begin{table}[H]
\caption{Process-related factors}
\label{tab:ProcessBenefits}
\centering
\scriptsize
\begin{tabular}
{@{}p{2cm}|p{10cm}@{}}
\hline 
\textbf{Characteristic}  & \textbf{Measures }\\ \hline 
\multirow {3}{*}{\pbox{2.2cm}{Process-related\\benefits}} 
& \textbf{Development independence between teams:} The migration from a monolithic architecture to a Microservices- oriented one changes the way in which the development team is organized. Typically, a development team is reorganized around the Microservices into small, cross-functional, and self-managed teams \cite{SP1}, \cite{SP3}, \cite{SP4}, \cite{SP9}, \cite{SP12}. \\    
& \textbf{Continuous delivery:} \cite{SP1} notes that the deployment in a Microservices environment is more complex, given the high number of deployment targets. Hence, the authors of \cite{SP1} suggest automating the deployment as much as possible. \\
& \textbf{Reusability:} Microservices are designed to be independent of their environment and other services \cite{SP11}. This facilitates their reusability.\\
\hline 
\end{tabular}
\end{table} 

\begin{table}[H]
\caption{Cost-related measures}
\label{tab:CostComparison}
\centering
\scriptsize
\begin{tabular}
{@{}p{2cm}|p{10cm}@{}}
\hline 
\textbf{Characteristic} & \textbf{Measure }\\ \hline 
\multirow {3}{*}{Personnel Cost} 
& \textbf{Development costs:} \cite{SP5} argues that Microservices reduce the development costs given that complex monolithic applications are broken down into a set of services that only provide a single functionality. Furthermore, most changes affect only one service instead of the whole system. \\
\hline 
\multirow {3}{*}{\pbox{2.2cm}{Infrastructure\\Cost}} 
& \textbf{Cost per hour:} Is a measure used to determine the infrastructure costs \cite{SP1}. According to the experiment done in \cite{SP3}, the Microservices architecture had lower infrastructure costs compared to monolithic designs. \\
& \textbf{Cost per million requests:} In comparison to cost per hour, this measure is based on the number of requests / usage of the infrastructure. \cite{SP3} uses the infrastructure costs of a million requests to compare different deployment scenarios. \\
\hline
\end{tabular}
\end{table} 


\subsubsection{Microservices Migration Effects}
The analysis of the characteristics and measures adopted in the empirical studies considered by the related works allowed us to classify a set of measures that are sensitive to variations when migrating to Microservices. The detailed mapping between the benefits and issues of each measure is reported in Table~\ref{tab:ProductBenefits}. 

\textbf{Product Characteristics}. Regarding product characteristics, performance is slightly reduced in Microservices. 

When considering the different measures adopted to measure \textbf{performance}, the usage of containers turned out to decrease performance. This is also confirmed by the higher number of CPU instructions needed to process a client request (path length), which is at least double that of monolithic systems and therefore results in high CPU utilization. 
However, the impact of the usage of different programming languages in different services is negligible. Even if different protocols have different interpreters for different languages, the computational time is comparable. 

When considering high \textbf{scalability} requirements, Microservices-based systems in general outperform monolithic systems in terms of resources needed. If a monolith is using all the resources, the way to handle more connections is to bring up a second instance. If a single microservice uses all the resources, only this service will need more instances.  Since scaling is easy and precise, this means only the necessary amount of resources is used. 
As a result, for the same amount of money spent on resources, microservices deliver more throughput.



\rf{The \textbf{availability} of Microservices-based system can be affected by the higher number  of moving parts compared to monolithic systems. However, differently than in monolithic systems, }
in the event of the failure of one Microservice, the remaining part of the system will still be available~\cite{Soldani2018}\cite{TaibiIEEECloud}. 
\rf{It is important to mention that Microservices do not provide high availability by default and maintaining high availability for microservices is not a simple task~\cite{Saquicela21}.} 
\rf{In order to investigate the practices  to maintain high-availability in Microservices-based systems, Marquez et al~\cite{marquez20} conducted a survey among 40 practitioners, highlighting 12 practices. Examples of these practices are ``Prevent remote procedure calls from waiting indefinitely for a response'' or ``Efficiently distributing incoming network traffic among groups of backend servers''.}

\textbf{Maintenance} is considered more expensive in the selected studies.  The selected studies agree that the  maintenance of a single Microservice is easier than maintaining the same feature in Microservices. However, testing is much more complex in Microservices ~\cite{SP7}, and the usage of different programming languages, the need for orchestration, and the overall system architecture increase the overall maintenance effort. Moreover, Microservices-based systems, should also take into account maintenance-related metrics between services, trying to reduce coupling and increase cohesion between services. For this purpose, the Structural Coupling (SC)~\cite{Panichella2021} might be used to      easily identify the coupling between services. 

\textbf{Cost-related measures}
The development effort of Microservices-based systems is reported to be higher than the development of monolithic systems~\cite{SP5}. However, \cite{SP1} and \cite{SP3}  report that infrastructure costs are usually lower for Microservices than for monolithic systems, mainly because of the possibility to scale only the service that need more resources instead of scaling the whole monolith.

\section{The Survey}
\label{Survey}

In this section, we present the survey on migration metrics that we performed as well as its results. We describe the research questions, the study design, the execution, and the data analysis, as well as the results of the survey.

\subsection{Goal and Research Questions}
\label{RQ}
We conducted a case study among developers and professionals in order to identify in practice which metrics they considered important before and after migration. 

Based on our goal, we derived the following research questions (RQs): 

\vspace{2mm}
\begin{tabular}{p{0.8cm}p{10cm}}
    \textbf{RQ1.} & Why did companies migrate to Microservices?  \\
    \textbf{RQ2.} & Which information/metrics was/were collected before and after the migration?\\
    \textbf{RQ3.} &  Which information/metrics was/were considered useful by the practitioners?
\end{tabular}

\vspace{2mm}
With \textbf{RQ1}, we aim to understand the main reasons why companies migrated to Microservices, i.e., to understand whether they considered only metrics related to these reasons or other aspects as well. For example, we expect that companies that migrate to increase velocity considered velocity as a metric, but we also expect them to consider other information not related to velocity, such as maintenance effort or deployment time. 

With \textbf{RQ2}, we want to understand the information/metrics that companies considered as decision factors for migrating to Microservices. However, we are also interested in understanding whether they also collected this information/these metrics during and after the development of Microservices-based systems.

With \textbf{RQ3,} we want to understand which information/metrics practitioners considered useful to collect the migration process, and which they did not collect but now believe they should have collected. 

\subsection{Study Design}
\label{StudyDesign}
The information was collected by means of a questionnaire composed of five sections, as described in the following:
\begin{itemize}
\item \textit{\textbf{Demographic information}}: In order to define the respondents' profile, we collected demographic background information. This information considered predominant roles and relative experience. We also collected company information such as application domain, organization’s size via number of employees, and number of employees in the respondents' own team.
\item \textit{\textbf{Project information}}: We collected the following information on the project migrated to Microservices: creation and migration dates of the project.

\item \textit{\textbf{Migration motivations} (RQ1)}: In this section, we collected information on the reasons for migrating to microservices. 

\item \textit{\textbf{Migration information/metrics} (RQ2)}: This section was composed of two main questions:  \begin{itemize}
    \item Which information/metrics were considered \textbf{before} the migration, to decide if migrate or not? 
    
    \item Which information/metrics were considered \textbf{after} the migration, to decide if migrate or not? 
  \end{itemize}


\item \textit{\textbf{Perceived usefulness of the collected information/metrics} (RQ3)}: 
In this section, we collected information on the usefulness of an assessment framework based on the metrics identified and ranked in the previous section. The goal was to understand whether the set of metrics could be useful for deciding whether to migrate a system or not in the future.  

This section was based on three questions: 
\begin{itemize}
\item Here we ask to rank how useful is each metric proposed in the Literature (Table~\ref{tab:ProductBenefits}) and mentioned by the interviewee to decide if migrate to microservices or not. The ranking is based on a 6-point Likert scale, where 1 means absolutely not and 6 absolutely. 
\item How easy are the factors and measures to collect and use? 
\item Which factor or measure is not easy to collect?
\item How useful is a possible discussion of the factors and measures reported in the previous questions before the migration? The ranking is based on a 6-point Likert scale, where 1 means absolutely not and 6 absolutely.
\item Do you think the factors or measures support a reasoned choice of migrating or not? (if not, please motivate)

\item Would you use this set of factors and measures in the future, in case of migration of other systems to Microservices? If not, please motivate.
\end{itemize}

\end{itemize}

The questionnaire adopted in the interviews is reported in \textit{Appendix}. 

\subsection{Study Execution}
\label{StudyExecution}
The survey was conducted over the course of five days, during the 19\textsuperscript{th} International Conference on Agile Processes in Software Engineering, and Extreme Programming (XP 2018). We interviewed a total of 52 practitioners. 
We selected only experienced participants that successfully developed the microservices-based system and deployed in production. We did not consider any profiles coming from academia, such as researchers or students. 

\subsection{Data Analysis}
\label{Dataanalysis}
Two authors manually produced a transcript of the answers of each interview and then provided a hierarchical set of codes from all the transcribed answers, applying the open coding methodology~\cite{Wuetherick2010}. The authors discussed and resolved coding discrepancies and then applied the axial coding methodology~\cite{Wuetherick2010}. 

Nominal data was analyzed by determining the proportion of responses in each category. Ordinal data, such as 5-point Likert scales, was not converted into numerical equivalents since using a conversion from ordinal to numerical data entails the risk that any subsequent analysis will yield misleading results if the equidistance between the values cannot be guaranteed. Moreover, analyzing each value of the scale allowed us to better identify the potential distribution of the answers.
Open questions were analyzed via open and selective coding~\cite{Wuetherick2010}. The answers were interpreted by extracting concrete sets of similar answers and grouping them based on their perceived similarity.

\subsection{Replication}
\label{Replication}
In order to allow replication and extension of our work, we prepared a replication package with the results obtained~\footnote{Raw data available at \url{https://figshare.com/s/cb8314fb66163d9fcdc9}. } The complete questionnaire is reported in Appendix.  

\subsection{Results}
\label{sec:survey_Results}
In this section, we will report the obtained results, including the demographic information regarding the respondents, information about the projects migrated to Microservices, and the answers to our research questions. 

\textbf{Demographic information}. The respondents were mainly working as developers (31 out of 52) and project managers (11 out of 52), as shown in Table~\ref{tab:Role}. The majority (23 out of 52) of them had between 2 and 5 years of experience in this role (Table~\ref{tab:Experience}).
Regarding company information, out of the 52 respondents, 10 worked in IT consultant companies, 6 in software houses, 8 in e-commerce, and 6 in banks. The remaining 9 respondents who provided an answer worked in different domains (Table~\ref{tab:Domain}).
The majority of the companies (15 out of 52 respondents) were small and medium-sized enterprises (SMEs) with a number of employees between 100 and 200, while 9 companies had less than 50 employees. We also interviewed people from 3 large companies with more than 300 employees (Table~\ref{tab:Organization}).
Regarding the team size, the vast majority of the teams had less than 50 members (33 out of 52 respondents). 14 teams had less than 10 members, 12 teams had between 10 and 20 members, and 7 teams had between 20 and 50 members. Only one team was composed of more than 50 members (Table~\ref{tab:Team}).

\begin{table}[H]
\centering
\begin{tabular}{ll}

\begin{minipage}{0.48\textwidth}
\begin{table}[H]
\caption{Role}
\label{tab:Role}
\centering
\footnotesize
\begin{tabular}
{@{}p{3cm}|p{2cm}@{}}
\hline 
\textbf{Role} & \textbf{\#Answers} \\ \hline 
Developer	&31\\ \hline 
Project Manager	&11\\ \hline 
Agile Coach	&2\\ \hline 
Architect	&2\\ \hline 
Upper Manager	&2\\ \hline 
Other	& 5\\ \hline 
\end{tabular}
\end{table}
\end{minipage}

& 

\begin{minipage}{0.48\textwidth}
\begin{table}[H]
\caption{Experience (in Years)}
\label{tab:Experience}
\centering
\footnotesize
\begin{tabular}
{@{}p{3.2cm}|p{2cm}@{}}
\hline 
\textbf{Experience in years} & \textbf{\# Answers} \\ \hline 
years $\leq$ 2  & 2 \\ \hline 
2 $<$ years $\leq$ 5  & 23 \\ \hline 
5 $<$ years $\leq$ 8  & 12 \\ \hline 
8 $<$ years $\leq$ 10  & 11 \\ \hline 
10 $<$ years $\leq$ 15  & 3 \\ \hline 
(no answer) & 1 \\ \hline
\end{tabular}
\end{table}
\end{minipage}
\\

\begin{minipage}{0.48\textwidth}
\begin{table}[H]
\caption{Organization Domain}
\label{tab:Domain}
\centering
\footnotesize
\begin{tabular}
{@{}p{3cm}|p{2cm}@{}}
\hline 
\textbf{Organiz. Domain} & \textbf{\# Answers} \\ \hline 
IT consultant  & 10 \\ \hline 
Banking & 6 \\ \hline 
Software house & 6 \\ \hline 
E-commerce  & 8 \\ \hline 
Other  & 9 \\ \hline 
(no answer) & 13 \\ \hline
\end{tabular}
\end{table}
\end{minipage}

&

\begin{minipage}{0.48\textwidth}
\begin{table}[H]
\caption{Team Size}
\label{tab:Team}
\centering
\footnotesize
\begin{tabular}
{@{}p{3cm}|p{2cm}@{}}
\hline 
\textbf{\# Team Members} & \textbf{\# Answers} \\ \hline 
\# $\leq$ 10  & 14 \\ \hline 
10 $<$ \# $\leq$ 20  & 12 \\ \hline 
20 $<$ \# $\leq$ 50  & 7 \\ \hline 
\# $>$ 50  & 1 \\ \hline 
(no answer) & 18 \\ \hline
\end{tabular}
\end{table}
\end{minipage}
\end{tabular}
\end{table}

\begin{table}[H]
\caption{Organization Size}
\label{tab:Organization}
\centering
\footnotesize
\begin{tabular}
{@{}p{6.3cm}|p{2cm}@{}}
\hline 
\textbf{\# Employees in Organization} & \textbf{\# Answers} \\ \hline 
\# organization employees $\leq$ 50  & 9 \\ \hline 
50 $<$ \# organization employees $\leq$ 100  & 0 \\ \hline 
100 $<$ \# organization employees $\leq$ 200  & 15 \\ \hline 
200 $<$ \# organization employees $\leq$ 300  & 3 \\ \hline 
\# organization employees $>$ 300  & 8 \\ \hline 

(no answer) & 19 \\ \hline
\end{tabular}
\end{table}

\textbf{Project information}
As for the project's age (Table~\ref{tab:ApplicationAge}), about 69\% of the respondents (36 out of 52) started the development less than 10 years ago, while 9 interviewees created the project between 10 and 15 years ago. Another 8 interviewees referred to projects with an age between 15 and 20 years, while 5 respondents started the development more than 20 years ago.
As for the migration to Microservices,    
23 respondents reported that the process started 2 years ago or less, while for 20 interviewees the process started between 2 and 4 years ago.

\begin{table}[H]
\caption{Application Age}
\label{tab:ApplicationAge}
\centering
\footnotesize
\begin{tabular}
{@{}p{3cm}|p{2cm}@{}}
\hline 
\textbf{Application Age} & \textbf{\# Answers} \\ \hline 
years $<$ 5             & 18 \\ \hline 
5 $<$ years $\leq$ 10   & 18 \\ \hline
10 $<$ years $\leq$ 15  & 9 \\ \hline 
15 $<$ years $\leq$ 20  & 3 \\ \hline 
years $>$ 20            & 5 \\ \hline 
\end{tabular}
\end{table}

\begin{table}[H]
\caption{Migration Time}
\label{tab:MigrationTime}
\centering
\footnotesize
\begin{tabular}
{@{}p{3cm}|p{2cm}@{}}
\hline 
\textbf{Migration Time} & \textbf{\# Answers} \\ \hline 
year $\leq$ 2  &  23 \\ \hline 
2 $<$ year $\leq$ 4  & 20 \\ \hline 
4 $<$ year   & 3 \\ \hline 
(no answer)  &  6 \\ \hline 
\end{tabular}
\end{table}

\subsubsection{Migration Motivations (RQ1)}

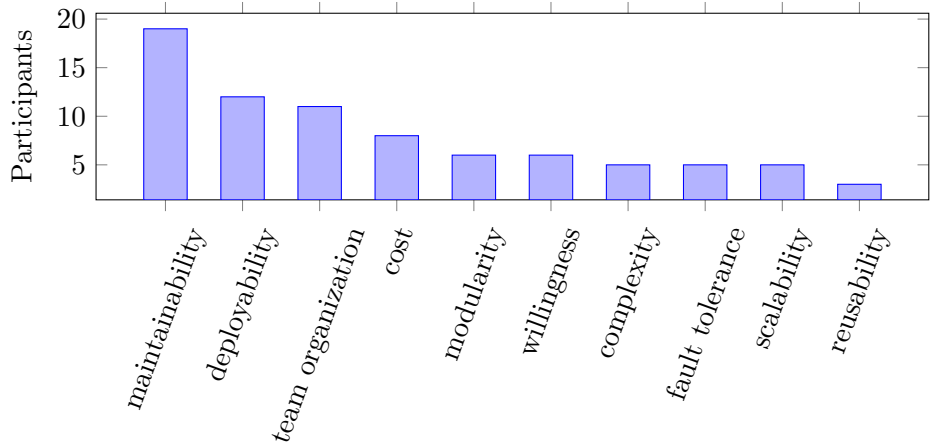
\begin{figure}[!ht]
\centering
\begin{tikzpicture}
\begin{axis}[
	height=1.6in,
    width=\linewidth,
    xtick={1,2,3,4,5,6,7,8,9,10},
    xticklabels={maintainability,deployability,team organization,cost, modularity, willingness, complexity, fault tolerance, scalability, reusability},
    ytick={0,5,10,15,20},
    ylabel=Participants,
    bar width=1.5em,
    xticklabel style={rotate=70},
	ybar 
]
\addplot coordinates {(1,19) (2,12) (3,11)(4,8)(5,6)(6,6)(7,5)(8,5)(9,5)(10,3) };
\end{axis}
\end{tikzpicture}
\caption{Migration motivations mentioned by more than three participants}
\label{fig:MigrationMotivations}
\end{figure}

In the answers to the question about the interviewees' motivation to migrate from their existing architecture to Microservices, a total of 97 reasons were mentioned. The open coding of the answers classified the 97 reasons into 22 motivations. In Figure~\ref{fig:MigrationMotivations}, all motivations that were mentioned three or more times are presented. The three main motivations are maintainability, deployability, and team organization.

The most commonly mentioned motivation was to improve the maintainability of the system (19 out of 97). They reported, among other things, that the maintenance of the existing system had become too expensive due to increased complexity, legacy technology, or size of the code base. 

Deployability was another important motivation for many interviewees (12 out of 97). They expected improved deployability of their system after the migration. The improvement they hoped to achieve with the migration was a reduction of the delivery times of the software itself as well as of updates. Moreover, some interviewees saw the migration as an important enabler for automated deployment (continuous deployment).

The third most frequently mentioned motivation was not related to expected technical effects of the migration but was organizational in nature, namely team organization (11 out of 97). With the migration to Microservices, the interviewees expected to improve the autonomy of teams, delegate the responsibility placed on teams, and reduce the need for synchronization between teams.

The remaining motivations like cost, modularity, willingness to adopt microservices because other companies are also adopting them, or the reduction of the overall system complexity seem to be motivations that are part of the three main motivations discussed above, or at least influence one of them. For example, complexity was often mentioned in combination with maintenance, or scalability together with team organization. Thus, it appears that these three motivations are the main overall motivations for the migration from monoliths to Microservices.

\subsubsection{Information/metrics collected before and after the migration (RQ2)}


We collected 46 different pieces of information/metrics, which were considered a total of 107 times by the interviewees \textbf{before} the migration to Microservices. The three most commonly mentioned ones were the number of bugs, complexity, and maintenance effort (see Table \ref{tab:metricsBeforeDuring}), followed by the velocity, and response time. Other five motivations were mentioned less frequently.

\begin{table}[H]
\caption{Information/metrics considered before the migration mentioned at least three times.}
\label{tab:metricsBeforeDuring}
\centering
\footnotesize
\begin{tabular}
{@{}p{4cm}|p{2cm}@{}}
\hline 
\textbf{Information/Metrics} & \textbf{\# Answers} \\ \hline 
Number of bugs  &  16 \\ \hline 
Complexity  & 11 \\ \hline 
Maintenance effort  & 10 \\ \hline 
Velocity  & 6  \\ \hline 
Response time & 6 \\ \hline
Lines of code & 3 \\ \hline
Performance & 3 \\ \hline
Extensibility & 3 \\ \hline
Change frequency & 3  \\ \hline
Scalability & 3 \\ \hline
\end{tabular}
\end{table}

Considering the information/metrics that collected \textbf{after} migration to Microservices, 26 clearly distinguishable types were identified that were mentioned a total of 66 times by the participants. Again, the number of bugs, complexity, and maintenance effort were the most frequently mentioned ones. (see Table \ref{tab:metricsAfter})

\begin{table}[H]
\caption{Information/metrics considered after the migration, mentioned at least three times.}
\label{tab:metricsAfter}
\centering
\footnotesize
\begin{tabular}
{@{}p{4cm}|p{2cm}@{}}
\hline 
\textbf{Information/Metrics} & \textbf{\# Answers} \\ \hline 
Number of bugs  &  12 \\ \hline 
Complexity  & 9 \\ \hline 
Maintenance effort  & 7 \\ \hline 
Velocity  & 5  \\ \hline 
Scalability & 5 \\ \hline
Memory consumption & 3 \\ \hline
Extensibility & 3 \\ \hline
\end{tabular}
\end{table}

As expected, the vast majority of the considered information/metrics was aimed at measuring characteristics related to the migration motivations. As maintainability was the most important reason to migrate to Microservices, maintainability-related metrics turned out to be the most important metrics considered before the migration. It is interesting to note that in some cases, companies collected this information before the migration but stopped collecting it during and after the migration (e.g., 4 interviewees out of 16 who had collected the number of bugs in their monolithic system did not collect the same information in the Microservices-based system).

The results suggest that the most important information needs remain the same from the start of the migration until its completion. Thus, there may be a set of migration information/metrics that is fundamentally important for the process of migration and that should be collected and measured throughout the migration.



\subsubsection{Information/metrics considered useful (RQ3)} 

In this section, we will report the results on the perceived usefulness of the metrics collected.

Asking the interviewees how \textbf{easy} they think it is \textbf{to collect} the factors and measures proposed, 41 answered that they considered them as easy, while 10 did not consider them as easy (one interviewee did not provide an answer to this question). While entering into the details of the metrics not easy to collect, only a limited number of interviewees mentioned some metrics as complex. 20 different metrics were reported, but only complexity was mentioned by six interviewees while four metrics (testability, response time, benchmark data, and availability) were considered as complex by only two interviewee and the remaining 15 metrics were mentioned only by one participant (see Table \ref{tab:measuresDifficult}).

\begin{table}[H]
\caption{Information/metrics not easy to collect mentioned by more than one interviewee.}
\label{tab:measuresDifficult}
\centering
\footnotesize
\begin{tabular}
{@{}p{3cm}|p{2cm}@{}}
\hline 
\textbf{Metric} & \textbf{\# Answers} \\ \hline 
Complexity  &  6 \\ \hline 
Testability  & 2 \\ \hline 
Response time & 2 \\ \hline 
Benchmark data & 2  \\ \hline 
Availability & 2 \\ \hline
\end{tabular}
\end{table}

Almost all interviewees categorized the usefulness of the metrics as very useful (24 out of 52) or extremely useful (25 out of 52). Table~\ref{tab:metricsusefulness} reports the medians for the usefulness of each metric reported by the interviewees.
Considering the \textbf{usefulness} of a possible discussion of the factors and measures reported in RQ2 before the migration, the majority of the interviewees considered it as very useful
 to understand the importance of the migration.
(see Table \ref{tab:measuresUsefulness}). Furthermore, all but three interviewees confirmed that they believe that the metrics support a rational choice on whether to migrate or not.

Finally, 65\% (34 out of 52) of the interviewees stated that they would consider the set of information/metrics proposed in the future, before migrating to microservice.

\begin{table}[H]
\caption{How useful did the interviewees consider each metric before migration.}
\label{tab:metricsusefulness}
\centering
\footnotesize
\begin{tabular}
{@{}p{10cm}|p{2cm}@{}}
\hline 
\textbf{Usefulness} & \textbf{Median} \\ \hline 
Response time (the time between sending a request and receiving the corresponding response)	&	4	 \\ \hline 
Cpu utilization (the percentage of time the cpu is not idle)	&	4	 \\ \hline 
Path length (the number of cpu instructions to process a client request)	&	4	 \\ \hline 
Waiting time (the time a service request spends in a waiting queue before it get processed)	&	4	 \\ \hline 
Impact of programming language (communication between microservices are network based)	&	4	 \\ \hline 
Usage of containers (the usage of containers can influence the performance, since they need additional computational time compared to monolithic applications deployed in a single container)	&	4	 \\ \hline 
Number of features per microservices	&	4.5	 \\ \hline 
Number of requests per minute or second (also referred as throughput or average latency)	&	5	 \\ \hline 
Downtime	&	5	 \\ \hline 
Mean time to recover (the mean time it takes to repair a failure and return back to operations)	&	5	 \\ \hline 
Mean time to failure (the mean time till the first failure)	&	5	 \\ \hline 
Testability	&	5	 \\ \hline 
Complexity	&	5	 \\ \hline 
Development independence between teams (the migration from a monolithic architecture to a microservice oriented changes the way in which the development team is organized)	&	5	 \\ \hline 
Continuous delivery	&	5	 \\ \hline 
Reusability	&	5	 \\ \hline 
Personnel cost (development cost)	&	4	 \\ \hline 
Infrastructure cost (cost per hour)	&	4	 \\ \hline 
Infrastructure cost (cost per million of requests)	&	4	 \\ \hline \hline

\multicolumn{2}{c}{Likert scale: 1-Absolutely not, 2-Little, 3-Just enough, 4-More than enough,} \\
\multicolumn{2}{c}{5-Very/a lot, 6-Extremely useful} \\ 

\end{tabular}
\end{table}

\begin{table}[H]
\caption{How useful did the interviewees consider discussion of the set of information/metrics before migration.}
\label{tab:measuresUsefulness}
\centering
\footnotesize
\begin{tabular}
{@{}p{3cm}|p{2cm}@{}}
\hline 
\textbf{Usefulness} & \textbf{\# Answers} \\ \hline 
Absolutely not &  0 \\ \hline 
Little & 0 \\ \hline 
Just enough & 1 \\ \hline 
More than enough & 2  \\ \hline 
Very/a lot & 24 \\ \hline
Absolutely & 25 \\ \hline
\end{tabular}
\end{table}

\section{The Assessment Framework}
\label{Framework}

In this section, we propose an evidence-based assessment framework based on the characteristics that should be considered before migration to identify and measure potential benefits and issues of the migration. The framework is evidence-based in the sense of evidence-based software engineering~\cite{kitchenham2004evidence} as it has not been derived based purely on subjective experience of the authors, but rigorously based on a systematic literature study and a survey.

The goal of the framework is to support companies in reasoning about the usefulness of migration and make decisions based on real facts and actual issues regarding their existing monolithic systems. 
The framework is not aimed at prescribing a specific decision, such as recommending to migrate based on a specific metric, but it is aimed at helping companies to not miss important aspects and to reason on the most complete set of information before deciding to migrate or not. The framework therefore has to be tailored to the specific contexts of companies that apply it.

Based on the results obtained in our survey (Section~\ref{Survey}), we grouped the different pieces of information and metrics into homogeneous categories, based on the classification proposed by the ISO/IEC 25010 standard~\cite{iso25010}. However, we also considered two extra categories not included in ISO/IEC 25010, which focus on product characteristics, namely cost and processes. 

The framework is applied in four steps: 
\begin{itemize}[leftmargin=25mm]
\item[Step 1] Motivation reasons identification
\item[Step 2] Metrics identification
\item[Step 3] Migration decisions
\item[Step 4] Migration 
\end{itemize}

In the next sub-sections, we will describe each of the four steps in detail. 

\subsection{Motivations reasons identification}

Before migrating to Microservices, companies should clarify why they are migrating and discuss their motivation. As highlighted by previous studies~\cite{TaibiIEEECloud}~\cite{Soldani2018}, companies migrate to Microservice for various reasons and often migrate to solve some issues that need to be solved differently. Moreover, sometimes the migration can have negative impacts, for instance when companies do not have enough expertise or only have a small team that cannot work on different independent projects.
The quality characteristics listed in Table~\ref{tab:Results} could be used as a checklist to determine whether there is some common problem in the system that the company intends to solve with the migration.

Based on the motivation, companies should reason - optimally including the whole team in the process - on whether the migration could be the solution to their problems or whether it could create more issues than benefits. If, for any reason, it is not possible to include the whole team in this discussion, we recommend including at least the project manager and a software architect, ideally with knowledge about Microservices.

In case the team still wants to migrate to Microservices after this initial discussion, it could start discussing how to collect the metrics (Step 2). 

\subsection{STEP 2 - Metrics Identification} 
In order to finalize the decision on whether or not to migrate to Microservices, teams should first analyze their existing monolithic system. 
The system should be analyzed by considering the metrics reported in Table~\ref{tab:Results}. 

We recommend starting by considering the information and metrics related to the motivation for the migration. 
However, for the sake of completeness, we recommend discussing the whole set of metrics. For example, if a team needs to migrate to Microservices because of maintenance issues, they should not only consider the block "maintenance" but should also consider the remaining metrics, since other related information such as the independence between teams (process-related) could still be very relevant for maintenance purposes. 

The list of metrics reported in Table~\ref{tab:Results} is not meant to be complete for each characteristic, but is rather to be used as a reference guide for companies to help them consider all possible aspects that could affect the migration.  
For example, a company's monolithic system might suffer from performance issues (characteristic ''Functional Suitability''). The analysis of the sub-characteristics will help them to reason about ''Overall performance'', but they could also consider whether it is a problem related to ''Time behavior'' by analyzing the metric ''Response time'' and also considering the other sub-characteristics listed. However, if the motivation of the performance issue is different, the company will also be able to reason about it.


\begin{table}[]
\centering
\footnotesize
\caption{The Proposed Assessment Framework}
\label{tab:Results}
\adjustbox{center}{
\begin{tabular}{p{2.5cm}|p{3cm}p{3.4cm}p{6cm}} \hline 
\textbf{Characteristic } &	\textbf{Sub-characteristic} 	&	\textbf{Meausure}	&	\textbf{Metric}	\\ \hline 
\rowcolor{gray10}\multicolumn{4}{l}{Functional suitability}	\\ \hline 
	&	Appropriateness & &  System requirements understadanbility	\\ \hline
\rowcolor{gray10}\multicolumn{4}{l}{Performance efficiency}			\\\hline 
	&	Overall	&		&		\\ \cline{2-4}
	&	Time behaviour	&		& Response time, throughput, ...		\\\cline{2-4}
	&	Resource utilization	&	&		Memory\rf{, disk space, nodes, ...} 	\\\cline{2-4}
	&	Compliance	&	Scalability	&		\\ \cline{2-4}
	&	Other	&	 &	\#Requests	\\\hline 
\rowcolor{gray10}\multicolumn{4}{l}{Reliability}		\\\hline
	&	Overall	&		& Mean Time to Failure	\\ 
		&		&		& Mean Time to Repair 	\\
			&		&		& Mean Time Between Failure 	\\\cline{2-4}
	&	Availability	&	     	& \%Availability \\
	&		&	     	& Mean Time Between Downtimes 		\\ \cline{2-4}
	&	Fault tolerance	&	&\#Bugs			\\
		&		&&	Code coverage			\\
	&		&Impact of failures&	 			\#Feature blocked, ... \\ \cline{2-4}
	&	Other	&\rf{Backups}& 			\\\hline
\rowcolor{gray10}\multicolumn{4}{l}{Maintainability}			\\ \hline 
	&	Overall	&		&		\\\cline{2-4}
	&	Modularity     	&	Code complexity	&		\\
	&		&	Adopted patterns	&		\\\cline{2-4}
                                                 	&	Reusability	&		&		\\\cline{2-4}
	&	Testability	&&	Code coverage			\\\cline{2-4}
	&	Analyzability	&	&\#Microservices			\\
	&		&	& Complexity (code, data, ...)		\\
		&		&	& Interactions between services		\\
	\cline{2-4}
	&	Modifiability	&	Code size	&	\#Lines of code, ...	\\
	&		&	Change frequency	&		\\
	&		&	Coupling	&		\\
	&		&	Service responsibilities	&		\\\cline{2-4}
	&	Changeability	&&	Extensibility			\\\hline 
\rowcolor{gray10}\multicolumn{4}{l}{Cost}		\\\hline 
	&	Overall	&& Development, testing, \rf{deployment}  				\\\cline{2-4}
	&	Infrastructure	&		& Cloud/On-Premise infrastructural costs		\\\cline{2-4}
	&	Effort	&		& Overall development		\\
	&		&		&	Testing, \rf{deployment}, maintenance, ...	\\	\hline
\rowcolor{gray10}\multicolumn{4}{l}{Process related}\\ \hline 
	&		&		&	Independence between teams \\
	&		&		&	\#User stories done per sprint \\
	&		&		&	Data management	\\
	&		&		&	Delivery time  \\
	&		&		&	Deployment frequency \\
    &		&		&	Feature priorities \\
    &		&		&	Roadmap	\\
	&		&		&	Service responsibilities \\
	&		&		&	Team alignment	\\
	&		&		&	Velocity (lead time/time to release)	\\\hline
\end{tabular}
}
\end{table}


\subsection{Migration Decisions}

After a thorough discussion of the collected metrics, the team can decide whether to migrate or not based on the results of the discussion performed in the previous step. 

For example, there will be cases where a company may decide not to migrate after all. If the company realizes that the reason for the low performance is due to the inefficient implementation of an algorithm, they might decide to implement it better. If the main issue is cost of maintenance and the company wants to migrate mainly to reduce this cost, they might think of better team allocation or reason about the root causes of the high costs, instead of migrating with the hope that the investment will enable them to save money.   


\subsection{Migration}
The team can then start the migration to Microservices. During this phase, we recommend that companies automate measurement of the relevant metrics and set up measurement tools to continuously collect relevant information as identified in Step 2.
\section{Discussion}
\label{Discussion}


In this section, we will discuss the implications of this work.


The results of our survey are in line with the characteristics identified by the related work. The vast majority of the interviewees migrated to Microservices in order to improve maintainability~\cite{Soldani2018}\cite{TaibiIEEECloud}. However, deployability, team organization (such as the independence between teams), and cost are also important characteristics mentioned frequently in the interviews and not considered as important by pevious work. Modularity, complexity, fault tolerance, scalability, and reusability were mentioned several times as well.

The proposed framework therefore covers characteristics and sub-characteristics that take the results of the survey into account and are aligned with the established ISO/IEC 25010 standard. The top-level characteristics are functional suitability, reliability, maintainability, cost, and process. The characteristics cover all the relevant sub-characteristics and metrics identified in the survey. For instance, modularity is a sub-characteristic of maintainability and scalability is a metric for performance efficiency.

Finally, the framework suggests concrete metrics for measuring the characteristics.
Given that all discussed characteristics are covered by metrics identified in the papers, the metrics can be used as an initial tool set to measure the main influencing factors for migrating a monolithic system to Microservices. Some characteristics are not easy to quantify, however. For instance, testability has effectiveness and efficiency aspects that can only be approximated by different metrics~\cite{garousi2018survey}, like the degree of coverage or the number of defects covered. The survey was used to confirm the metrics found and to identify additional ones. The metrics most commonly mentioned in the survey are the number of bugs, complexity, and maintenance effort. It turns out that for the characteristics that are most relevant for migration, these metrics are also mentioned more often than for other characteristics. Maintainability is mentioned as the most important reason for migration, and maintainability-related metrics are also highlighted as the most important metrics. 

In our study, we discovered that practitioners often do not properly measure their product, process, and cost before migrating to Microservices and realize only later (during or after the migration) that relevant information is missing. Our proposed assessment framework should not only help to identify the most relevant characteristics and metrics for migration, but also make professionals aware of the importance of measurement before, during, and after migration to Microservices. In addition, there has not been a clear understanding what to measure before migrating to Microservices. Our proposed assessment framework intends to fill this gap. However, evaluation and refinement of the framework in industrial case studies is required as part of future work.
\section{Threats to Validity}
\label{Threats}

We applied the structure suggested by Yin~\cite{YinCaseStudies2009} to report threats to the validity of this study and measures for mitigating them. We report internal validity, external validity, construct validity, and reliability. As we performed a mixed-methods approach comprising a Systematic Mapping Study and a survey, we will identify in this section different threats to validity regarding both parts of our study.

\subsection{Threats to Validity regarding the Survey}

\textbf{Internal Validity}. One limitation that is always a part of survey research is that surveys can only reveal the perceptions of the respondents which might not fully represent reality. However, our analysis was performed by means of semi-structured interviews, which gave the interviewers the possibility to request additional information regarding unclear or imprecise statements by the respondents. The responses were analyzed and quality-checked by a team of four researchers.

\textbf{External Validity}. Overall, a total of 52 practitioners were interviewed at the 19\textsuperscript{th} International Conference on Agile Processes in Software Engineering, and Extreme Programming (XP 2018). We considered only experienced respondents and did not accept any interviewees with an academic background. XP 2018 covers a broad range of participants from different domains who are interested in Microservices and the migration to Microservices. We therefore think that threats to external validity are reasonable. However, additional responses should be collected in the future.

The questions are aligned with standard terminology and cover the most relevant characteristics and metrics. In addition, the survey was conducted in interviews, which allowed both the interviewees and the interviewer to ask questions if something was unclear. 

\textbf{Reliability}. The survey design, its execution, and the analysis followed a strict protocol, which allows replication of the survey. However, the open questions were analyzed qualitatively, which is always subjective to some extent, but the resulting codes were documented.

\section{Conclusion}
\label{Conclusion}

In this paper, we proposed an assessment framework to support companies in reasoning on the usefulness of the migration to Microservices. 

We identified a set of characteristics and metrics that companies should discuss when they consider migrating to Microservices. 
The identification of these characteristics was performed by means of an industrial survey, where we interviewed 52 practitioners with experience in developing Microservices. The interviews were based on a questionnaire in which we asked the respondents to identify which metrics and characteristics had been adopted when they migrated to Microservices, which of these were useful, and which had not been adopted but should have been. The metrics were collected by means of open questions so as to avoid any bias of the results due to a set of predefined answers. After the open questions, we also asked the practitioners to check whether they had also collected some of the metrics proposed in the literature, and whether they believed it would have been useful to collect them. 

The result of this work is an assessment framework that can support companies in discussing whether it is necessary for them to migrate or not. The framework will help them avoid migration if it is not necessary, especially when they might get better results by refactoring their monolithic system or re-structuring their internal organization. 

Future work include the validation of the framework in industrial settings, and the identification of a set of automatically applicable measure, that could easily provide a set of meaningful information, reducing the subjectivity of the decisions. Another interesting future direction is the extension of this framework for different cloud-native technologies, including serverless~\cite{Nupponen2020}\cite{Taibi2021IEEE} and Micro-Frontends~\cite{Peltonen2021}. It would be interesting to investigate frameworks to enable practitioners to understand when it is beneficial to migrate from monolithic to serverless functions, and in particular, which serverless pattern to adopt~\cite{TaibiCLOSER2020} to create microservices based on serverless functions without decreasing productivity or increase technical debt~\cite{Lenarduzzi2021IEEE}
\section*{Acknowledgements}

This work was partially supported by the Austrian Science Fund (FWF): I 4701-N and by the Federal Ministry for Climate Action, Environment, Energy, Mobility, Innovation and Technology (BMK), the Federal Ministry for Digital and Economic Affairs (BMDW), and the Province of Upper Austria in the frame of the COMET - Competence Centers for Excellent Technologies Programme managed by Austrian Research Promotion Agency FFG.

\section*{References}
\bibliographystyle{model1-num-names}
\bibliography{sample.bib}

\section*{Appendix: The Survey}
I this Section we report the questionnaire adopted in the interviews. 

\vspace{5mm}
\noindent\textbf{Demographic information}
\begin{itemize}
    \item Company name
    \itemsep0em
    \item Respondent name
    \item Respondent email address
    \item Role in the organization
  \begin{itemize}  
    \item Upper Manager        
    \item Manager      
    \item Developer
    \item Other
  \end{itemize}
\item How many years have you spent in your role? 
\item Number of employees of your team
\item Number of employees of your organization
\item Organization’s domain(s)
\end{itemize}

\noindent\textbf{Project Information }
\begin{itemize}
\item Which microservices-based application is your company developing? 
\item When was the application first created? 
\item When did your company decide to migrate to microservices? 
\end{itemize}

\noindent\textbf{Migration Motivations }
\begin{itemize}
\item Why did your company decide to migrate? 
\end{itemize}

\noindent\textbf{Migration Information/Metrics }
\begin{itemize}
\item Which information/metrics were considered \textbf{before} the migration? 
\item Which information/metrics were considered \textbf{after} the migration? 
\end{itemize}

\noindent\textbf{Perceived usefulness of the collected Information/Metrics}
\begin{itemize}
\item We developed a set of factors and measures to support companies in evaluating the migration to microservices before they start, based on the assessment of a set of information to support them in reasoning about the needs of migrating. 
\item Which of the following information/metrics do you consider  useful to collect and discuss before the migration?

\begin{table}[H]
\footnotesize
\begin{tabular}{p{7.5cm}|p{0.5cm}|p{0.5cm}|p{0.5cm}|p{0.5cm}|p{0.5cm}|p{0.5cm}} \hline
& \rotatebox{90}{absolutely not} & \rotatebox{90}{little} & \rotatebox{90}{just enough} & \rotatebox{90}{more than enough } & \rotatebox{90}{very/a lot} & \rotatebox{90}{absolutely} \\

\textbf{Scalability/Performance } &                &        &             &                  &            &            \\ \hline

-  Response time                                                                                                                                                     &                &        &             &                  &            &            \\
\tiny(The time between sending a request and receiving the corresponding response)                                                                                             &                &        &             &                  &            &            \\
-  CPU utilization                                                                                                                                                    &                &        &             &                  &            &            \\
\tiny(The percentage of time the CPU is not idle)                                                                                                                              &                &        &             &                  &            &            \\
-  Path length                                                                                                                                                        &                &        &             &                  &            &            \\
\tiny(The number of CPU instructions to process a client request)                                                                                                              &                &        &             &                  &            &            \\
-  Waiting time                                                                                                                                                       &                &        &             &                  &            &            \\
\tiny(The time a service request spends in a waiting queue before it get processed)                                                                                            &                &        &             &                  &            &            \\
-  Impact of programming language                                                                                                                                     &                &        &             &                  &            &            \\
\tiny(Communication between microservices are network based)                                                                                                                   &                &        &             &                  &            &            \\
-  Usage of containers                                                                                                                                                &                &        &             &                  &            &            \\
\tiny(The usage of containers can influence the performance, since they need additional computational time compared to monolithic applications deployed in a single container) &                &        &             &                  &            &            \\
-  Number of features per microservices                                                                                                                               &                &        &             &                  &            &            \\
-  Number of requests per minute or second                                                                                                                            &                &        &             &                  &            &            \\
\tiny(Also referred as throughput or average latency)                                                                                                                          &                &        &             &                  &            &            \\ \hline 
\textbf{Availability  }                                                                                                                                                            &                &        &             &                  &            &            \\
-  Downtime                                                                                                                                                           &                &        &             &                  &            &            \\
-  Mean time to recover                                                                                                                                               &                &        &             &                  &            &            \\
\tiny(The mean time it takes to repair a failure and return back to operations)                                                                                                &                &        &             &                  &            &            \\
-  Mean time to failure                                                                                                                                               &                &        &             &                  &            &            \\
\tiny(The mean time till the first failure)                                                                                                                                    &                &        &             &                  &            &            \\ \hline 
\textbf{Maintenance }                                                                                                                                                              &                &        &             &                  &            &            \\
-  Testability                                                                                                                                                        &                &        &             &                  &            &            \\
-  Complexity                                                                                                                                                         &                &        &             &                  &            &            \\ \hline 
\textbf{Process related benefits}                                                                                                                                                  &                &        &             &                  &            &            \\
-  Development independence between teams                                                                                                                             &                &        &             &                  &            &            \\
\tiny(The migration from a monolithic architecture to a microservice oriented changes the way in which the development team is organized)                                      &                &        &             &                  &            &            \\
-  Continuous delivery                                                                                                                                                &                &        &             &                  &            &            \\
-  Reusability                                                                                                                                                        &                &        &             &                  &            &            \\ \hline
\textbf{Personnel Cost }                                                                                                                                                           &                &        &             &                  &            &            \\
-  Development Cost                                                                                                                                                   &                &        &             &                  &            &            \\ \hline 
\textbf{Infrastructure Cost   }                                                                                                                                                    &                &        &             &                  &            &            \\
-  Cost per hour                                                                                                                                                      &                &        &             &                  &            &            \\
-  Cost per million of requests                                                                                                                                       &                &        &             &                  &            &            \\ \hline 
\textbf{Which other factors or measures should be considered?} (please list and rank them)  &        &             &                  &            &             \\ \hline 
\end{tabular}
\end{table}

\item How useful would you consider a discussion of the previous information before migration? 
\item Do you think the factors or measures support a reasoned choice of migrating or not? (if not, please motivate)
\item How easy is the set of factors and measures to collect and use? 
\item Is there any measure that is not easy to collect? 
\item Would you use this set of factors and measures in the future, in case of migration of other systems to microservices? If not, please motivate. 
\end{itemize}

\end{document}